# *SHIELD*: *S*ocial sensing and *H*elp *I*n *E*mergency using mobi*L*e *D*evices

Gautam S. Thakur, Mukul Sharma, Ahmed Helmy

*Abstract*—**School and College campuses face a perceived threat of violent crimes and require a realistic plan against unpredictable emergencies and disasters. Existing emergency systems (e.g., 911, campus-wide alerts) are quite useful, but provide delayed response (often tens of minutes) and do not utilize proximity or locality. There is a need to augment such systems with proximity-based systems for more immediate response to attempt to prevent and deter crime. In this paper we propose *SHIELD*, an on-campus emergency rescue and alert management service. It is a fully distributed infrastructure-less platform based on proximity-enabled trust and cooperation. It relies on localized responses, sent using Bluetooth and/or WiFi on the fly to achieve minimal response time and maximal availability thereby augmenting the traditional notion of emergency services. Analysis of campus crime statistics and WLAN traces surprisingly show a strong positive correlation (over 55%) between on-campus crime statistics and spatio-temporal density distribution of on-campus mobile users. This result provides a motivation to develop such platform and points to the promise in reducing crime incidences. We also show an implementation of a prototype application to be used in such scenarios.**

## I. Introduction

The current emergency, alert and public safety systems take centralized approaches and do not tap any available local rescue service. For example, 911 and Emergency BlueTowers connects to a centralized Public Safety Answering Point, which then send emergency rescuers at the crime site. On average, it takes 10-20 minutes to arrive. Also, they require pre-established links and infrastructure, which may not be available everywhere, especially in areas affected by earthquake disaster and floods.

On the other hand, decentralized and distributed approaches of small handheld devices with short wavelength communication (Bluetooth, WiFi) give new dimension to express human activities never seen before and can be exploited for personal safety and rescue. They helped to realize great potential of service localization, proximity, participatory sensing and message relaying in multitude of ways, for example: inferring shared interest[8] and friendship networks[6], identifying social structure and; human behavior based message forwarding[7]. In a novel way, here we extend their underlying capabilities to augment current emergency rescue and alert response management systems. We propose ideas to develop: (1) trust from mobility driven user encounters (2) context aware signaling and service localization of historical crime log statistics, all as measures to provide (3) a preemptive response in averting the possibility of incident occurring, in a system we call SHIELD. As a reaction to minimize the average response time of an already occurred event, SHIELD maximizes the use of available local help in the vicinity of incidence.

However, the main challenge is not just to provide these services for a favorable performance; instead the system should also be efficient and accurate to transmit the signals in time and only to the trusted entities of the network. This requirement represents an additional constraint on the design and functionality of the system to maintain privacy and trust whilst ensuring reliable communication. Thus, to achieve operational independence and robustness in the process of distress signaling, SHIELD should also provide a set of guidelines to establish them and increase the cooperation in the network, hence regulating the flow of information/message in a controlled manner.

To augment upon the existing facilities and focus on these challenges, in the ensuing text we discuss a novel approach to develop trust and cooperation in the network based on (1) Number of Bluetooth encounters (2) Duration of Bluetooth encounters. Then, we propose a comprehensive trust model built on these and other contextual features. The trust model plays a vital role in privacy preservation of mobile users and sought to increase cooperation inside the network. We also propose a context aware energy efficient protocol that takes input from trust model and historical crime log statistics. Keeping in mind the limited resource of mobile devices, the protocol is adaptive and adjusts the parameter setting to better serve the nature of emergency and alert scenario. Finally, the proposed system can easily augment existing services (like 911) and bridge the gap to cater available localized services of first responder as quickly as possible.

In all, our design goals to develop such infrastructure-less distributed system includes: (1) Maximum availability, (2) Minimum response time, (3) Reliability of communication via network trust generation, (4) Scalability and (5) Cross-platform functionality to mitigate response irrespective of the device manufacturers.

## II. Related Work

### A. *Emergency and Rescue Systems in General:*

As mentioned, most of the existing systems are either centrally controlled or require third party support (Cell Towers etc). For example, university campuses deploy BlueTowers and standard text messaging systems like CampusED, e2Campus,

Panic n' Poke to alert students and faculties. However, they have some shortcomings: BlueTowers are not available everywhere and the SMS text service are expensive, passive and incur lot of resources. Usually, these SMS are sent in thousands, which overload the central system and affect other voice-data services with delayed throughput and unacceptable percentage loss of total messages sent. Instead targeting affected users; it amasses the whole subscribed community (No localization of emergency), which may create an unnecessary panic. The affected mobile users lack any network driven trust and cooperation, but need to coordinate on their level. Finally, they cannot be used in situations of disasters, earthquake where infrastructure collapses.

### *B. Other Approaches:*

The development of a robust and responsive system is critical to emergency management. Several prototypes have been proposed in the past. The authors in[14] proposed a dynamic data driven application framework that uses wireless call data to measure the abnormal movement patterns in the population. The need for a reliable communication and interoperability challenges among rescue teams from technological, sociological and organizational point of view are discussed in [9]. The barriers to technology adaption in emergency management and user capabilities are discussed in [10][11], which gives deployment level intricacies in such systems. Finally, some real time test beds and simulation scenario are modeled in [12][13], to help develop a system that can actually react in reality.

### *C. Understanding User behavioral patterns:*

Currently, numerous attempts are being made to understand user behavioral patterns from machine-sensed measurements[4][5]. They try to discover mobile users' social structures, periodic routines, mobility demographics and spatio-temporal mobility profiles. A detailed study on various expects of human behavioral patterns are done in [15][16][17]. On the same lines, authors in [1][2] investigated the social structures, community formation and derived expressions for the cooperation in the network based on similarity and density distribution. All these work motivates us to generate a framework that uses behavioral patterns in the context of developing trust and cooperation from multi-sensing capability of handheld devices.

### *D. Using Human Mobility as a Communication Paradigm:*

To uncover user behavioral patterns is not enough; we need some compelling reasons to develop a communication paradigm based on these patterns. An important work in[7] has proposed a mobility protocol that uses human behavior to transfer messages. In [18][19], authors considered the impact of mobility in designing communication protocols and provided ways to develop an effective communication system. These rationales gave an important motivation that a system can be developed from user behavioral patterns. It can use those characteristics features as a medium to establish a secure and timely communication in DTN like environment. Next, we discuss SHIELD architecture.

## III. SHIELD: Rationale and Architectural Overview

In this section, we discuss the SHIELD architecture as shown in Figure 1. Here, we assume mobile users are carrying handheld devices equipped with RF-communication capability. The main components of the SHIELD are:

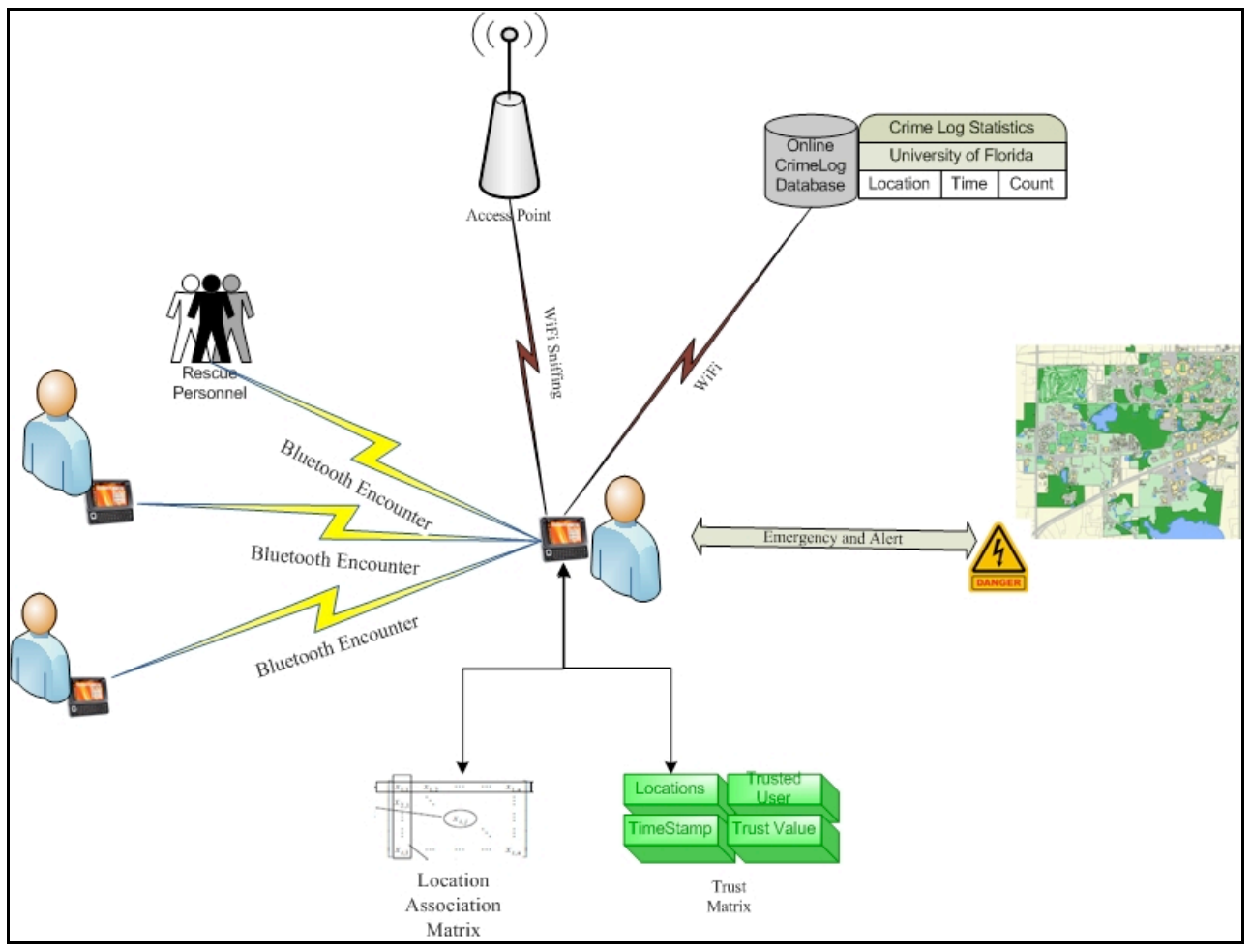

**Figure 1: The SHIELD Architecture**

### *A. The Encounter and Duration Matrix:*

The encounters are the discovery of the Bluetooth devices present in the vicinity. Initially, we build two matrices: 1) An Encounter matrix that contains the number of encounters with other users. 2) A Duration matrix that contains the duration of these encounters with other users. We also record the timestamp and the location of encounter. The location is derived from the access point sniffing and used to co-locate a user with the incidence location and time.

### *B. The Trust Matrix:*

The trust model discussed later uses these encounters and builds a wrapper of trust for the mobile host, in form of a trust matrix. It first maps encounters in spatio-temporal dimension and then assigns them into various classes of trust that identifies other users influence in emergency rescue and alert. From the analysis of real traces and human surveys, we found that large number of encounters and longer encounter duration belong to known persons like friends, colleagues and spouse.

### *C. Advisory Sub-System:*

To provide an optimal level of safety and self-preparation, we analyzed historical on-campus crime logs. We created an online database of these crimes statistics and ranked various campus location. The location ranking and vulnerability assessment is done using time and nature of the crime. The system gives a statutory threat warning to the mobile users whilst visiting a location at a particular time. The location is derived from the nearest Access Point. For example, the system flashes a cautionary signal to the students passing a parking garage at nighttime, if it has some recent crime history. As shown in the Figure 1, this data may be stored on the device, or accessed from a server using a WiFi connection.

### *D. Context-aware Adaptive Protocol:*

We introduce a context-aware adaptive protocol to complement trust model and advisory sub-system. Its main task is then to perform efficient routing and transmission of the distress signal. For example, during a critical time like passing a parking garage, the protocol increases Bluetooth scanning frequency to identify nearby trusted devices and notify them of its existence. However, in normal operations it tries to save resources (battery power etc), by reducing scanning frequency.

### *E. Distress Signaling:*

In emergency situations the mobile user can use all available modes of communication to let nearby trusted nodes know of the situation. A user can select classes of trust to send the distress signal (automatically) based on their availability and also to a category of individuals who provide specialized services like doctors, security and rescue personnel, nighttime vigil guards etc. In the following text, we describe the details.

In section IV, we discuss mobility measurements used to build encounter matrices, subsequently section V explains the trust model. The underlying context-aware adaptive protocol is illustrated in section VI. In section VII we develop the application prototype. Results and analysis to measure Bluetooth transfer performance and correlation statistics are mentioned in section VIII. Finally we conclude the paper in section IX with future direction.

## IV. TRACE ANALYSIS

An important aspect in Ad Hoc Network research is the careful logging of mobility traces and empirical ways to understand large systems. In the past few years, we saw a significant effort by several universities[4][5] to collect large-scale measurement that logs Bluetooth encounters and WLAN users' network usage spatio-temporal information. The *TRACE* framework as mentioned in[1] helps to further refine and generate encounter matrices. We use it to help us understand mobile users encounter patterns, spatio-temporal preferences and henceforth as a platform to develop a system that is built on mutual trust and cooperation to face emergency and critical situations. The table below shows the measurements used for our purpose of study.

| Type of Measurements | Duration | Statistics |
|---|---|---|
| Bluetooth Encounters | Fall 2009 | 135 Users with Nokia N810 |
| WLAN Usage | Fall 2008 | 12000 User |
| Crime Log | 1998-2010 | 17510 cases |

The above measurements are collected from the main university campus of University of Florida. These were collected from 135 students in Fall 2009 on Nokia N810s equipped with Bluetooth and WiFi Access Point sniffers. To understand usage, we also used WiFi measurements from Access Point (WLAN) connections. Another dataset is the crime log received from Police department of University of Florida, which contains a list of ten years of on-campus occurred incidences detailing the type, time, date and location of the crime.

## V. TRUST MODEL

Today mobile users frequently carry handheld (e.g. iPhones) devices that can be used to reflect their personality. Using the multi-sensor capability of these devices (e.g. Bluetooth, GPS and WiFi sniffing), we can accurately capture vital vicinity statistics like frequency and duration of time spent at particular locations. From the work done in [2][3][16] we know that statistical analysis of such historical logging of information pre-dominantly shows the mobile user behavioral pattern has location visiting preferences, periodic reappearances and preferential attachments.

Another perspective in the study of behavioral patterns is the analysis of similarity that helps develop inter-connection between mobile users. In this regard, a fundamental work is done in [20][21], which provides significant evidences of socio-demographic, spatio-temporal regularity and social structures as a basis to develop homophilous relations and propinquity among users. It also shows people who know each other, form a cohesive cluster with a small average shortest path length, and a large clustering coefficient. Using this rationale, we conducted an experiment to analyze user encounter patterns in University of Florida campus. For a period of ten weeks in Fall 2009, we distributed Nokia N810 and OpenMoko to 135 students. The devices were equipped to sniff nearby active Bluetooth devices in a range up to ~50 meters, localized by WiFi Access Point information. The analysis of WiFi and Bluetooth measurements provide an important result that large number and duration of meetings belong to users who know each other very well (validated by the students carrying out the experiments), which is shown in the Figure 2. The curves decrease sharply for the mobile users with known faces to the completely strangers.

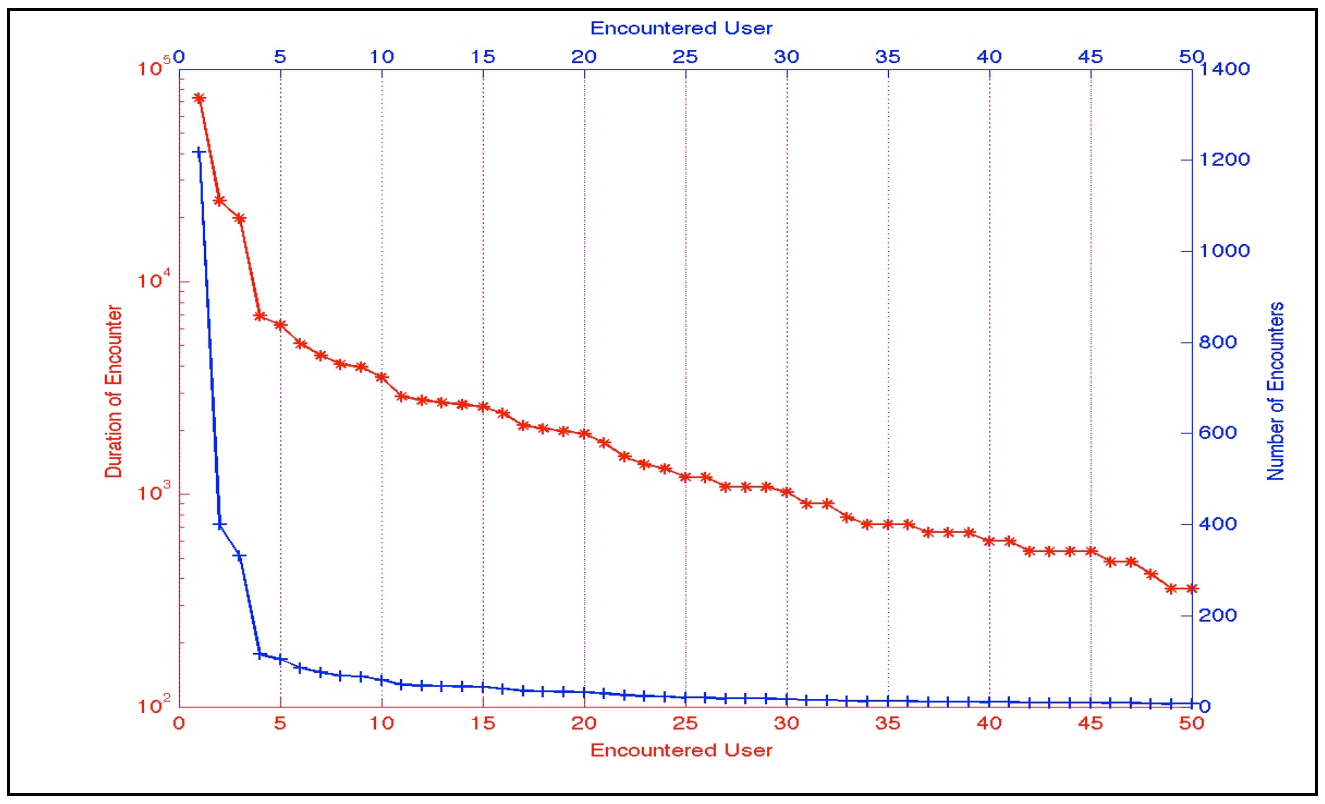

**Figure 2: A distribution shows mobile users who know one another have high frequency and large duration of encounters as compared to the strangers (right most section of both the curves).**

As explained in [22][23], characteristic similarity brings people of the same nature together. Thus, the information flows relevant to one mobile user more likely to be of the interest to another mobile user of same circle. Conversely, the same circle can be called for in the event of emergency and alert. Finally, the way to develop such a circle of trust and friendship can be derived from the mobile encounters. Using these results we can say that known mobile users with *frequent encounters* and *large meeting duration* most effectively are the ones whom can be trusted at first place. Furthermore, a co-operation network based on these two metrics can be developed and very well be used in the event of emergency and alert management. The formation of trust and cooperation between two mobile users $i$ and $j$ is defined that takes into consideration two factors:

### A. *Number of Encounters $f(i,j)$:*

We define the number of encounters as the frequency of encounter (i.e., coming within radio range) between two mobile users and are the number of repeated meetings per unit time. The meeting between two mobile users $i$ and $j$ is defined as

$$\delta(i,j) = \begin{cases} 1 & \text{if } i \text{ and } j \text{ meet} \\ 0 & \text{otherwise} \end{cases}$$

The number of encounters is thus defined as possible meetings over an evenly distributed set of continuous time interval of size $n$ as

$$f(i,j) = \sum_{i,j=1}^{n} \delta(i,j)$$

### B. *Duration of Encounters $D(i,j)$:*

We define duration of encounter as the amount of time spent by mobile users together. While the number of encounters provides an important criterion to quantify the active mobility of users in the network, it does not provide a minimum threshold time required to establish a connection and successfully transfer the messages. For example, say two mobile users often meet, but only for a fraction of a minute, despite a successful encounter, it is impossible for them to communicate effectively. Duration of encounters provides the requisite stability factor in a dynamic network environment. Qualitatively, it also defines the closeness between two mobile users, as a dimension to measure trustworthiness and an expected level of cooperation in the emergency and alert situation. From Figure 2, we see that a large number of encounters frequently entail longer meeting time with same set of users. Thus, we define the duration of encounter between two mobile user nodes $i$ and $j$ as:

$$D(i,j) = \sum_{i,j=1}^{n} d(\delta(i,j))$$

Where $d(\delta(i,j))$, is the individual duration of meeting between $i$ and $j$ while they encounter. These two metrics provide a foundation that led us to implement a comprehensive trust model. It effectively optimizes the distress signal transmission with trusted nodes. The trust model uses a rule-based classifier that recognize Bluetooth encounter and assigns them into various classes of trust. These classes of trust define the social proximity of a user in emergency and alert situations. The rule base classifier consists of a Rule set $R = \{r_1, r_2, r_3, \ldots, r_k\}$ such that each classification rule is of form:

$$r_j : (\text{Encounter, Condition}) = C_l(y)$$

Each rule consists of a condition statement that defines attributes pertinent to the encounter. The term $C_l(y)$ shows that node $y$ is assigned to class $C_l \in C$, such that $C = \{C_1, C_2, C_3, \ldots, C_m\}$. These rules may not be mutually exclusive and sometimes more than one rule can apply to an encounter. Following condition statements are used to built classifier from environment sensed emergent properties:

1) Location and vicinity information of Bluetooth encounter.
2) Tags that define the level of trust with an encountered device. These tags are similar to ranks and status quo of a person, i.e. doctors, security personnel.
3) Duration, frequency and clock time of the encounter.
4) Activity based encounters, which describes the circumstances when Bluetooth encountered happened.

This classifier is easy to interpret and can be incrementally built on the existing rules. Also, these rules help better express the social fuzziness that exists in the environment-sensed data.

## VI. PROTOCOL DESIGN

The protocol stack defines a process to send distress signals to a few trusted nodes. It achieves its goal to send a successful distress signal by managing human activities, sensing the emergent properties from the location of incidence, data of historical events and then intelligently choosing the most effective form of available communication.

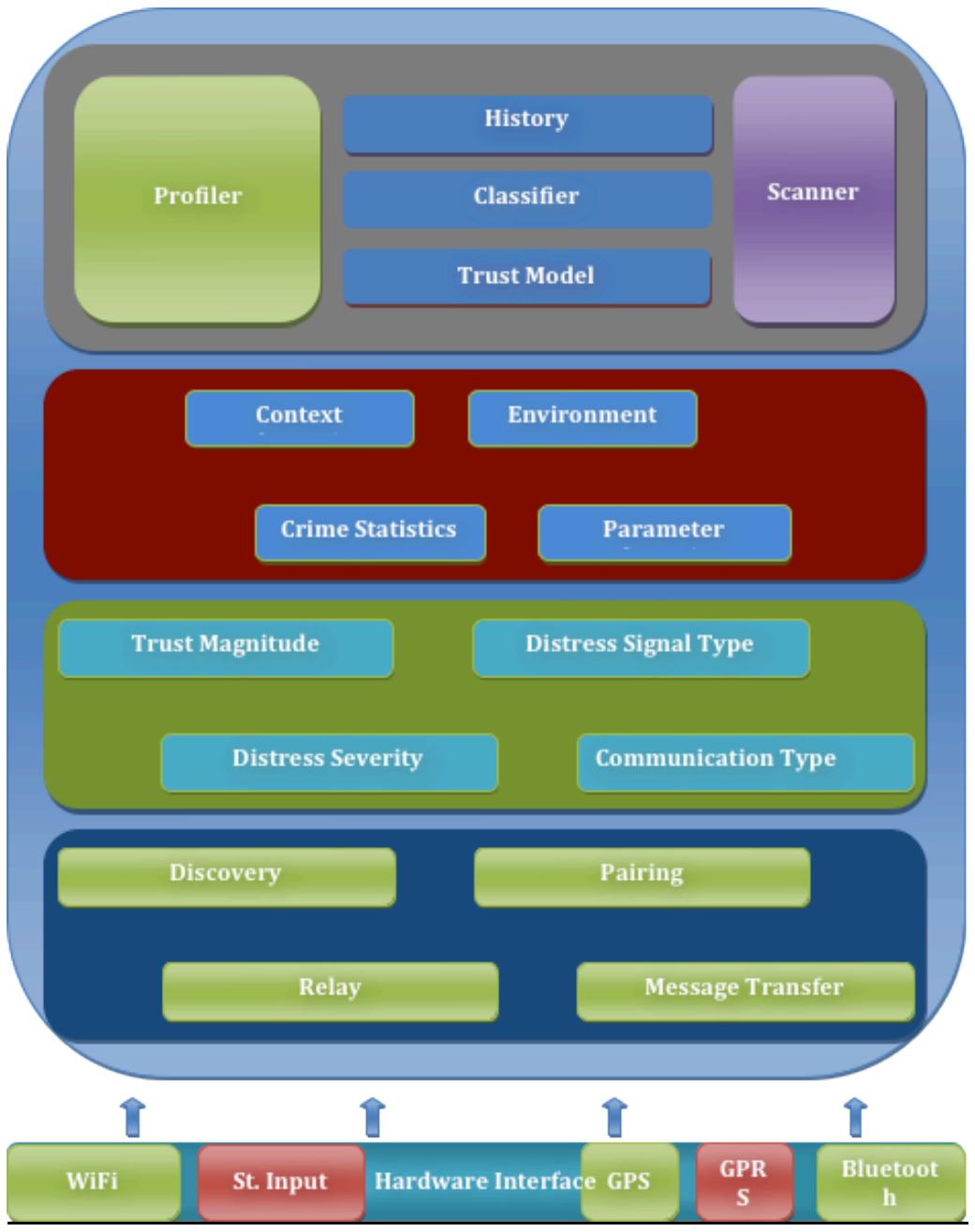

**Figure 3: Protocol Architecture**

The protocol stack is shown in the Figure 3 is divided into four main components. Each component of the protocol is described in more detail below:

### A. *The Scan Engine:*

The top component contains scanner and profiler for mobile user encounters. The trust model generates a user's behavioral profile and aggregates the classification of its Bluetooth measurements into the encounter matrix form. The device sensor provides detail on the historical crime log, location of the incident, duration & time to maximize efficiency of signal.

### B. *Protocol Adaption:*

To provide optimal level of services and to ensure the limitation imposed by mobile device, our protocol adapts to environment by selectively changing the parameter space based on environment sensed input and crime log statistics.

### *C. Distress Signal Communication:*

This component is responsible for deciding the level of trust. Based on the type of incident, the trust magnitude and severity is decided before sending the distress signal. We define various trust magnitudes: Friends, Strangers, Acquaintances, Tagged data etc. A distress signal can be bifurcated into emergencies and alerts. An *emergency* situation can be burglary, heart attack, while *alerts* might involve hurricane warning, earthquakes. Based on the panic, a victim can assign the level of severity to the distress signal. The application decides the life span of the message, number of hops, type of forwarding method etc.

### *D. Discovering and Pairing:*

The lower most component is responsible for the message transmission to other mobile user devices and also performs important operations like pairing, discovering and relaying to other devices. This module is attached to the hardware, which can use any of the available communication type to send the distress signal.

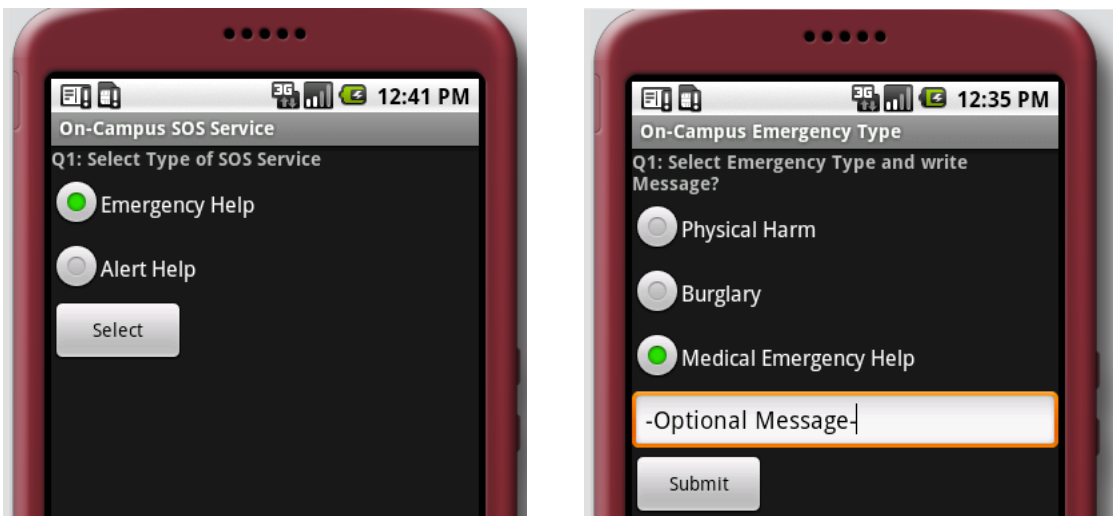

**Figure 4: Android Simulator showing SHIELD implementation**

## VII. APPLICATION PROTOTYPE

We developed a prototype application on Android Simulator that uses SHIELD architecture. As show in Figure 4, this prototype can use Bluetooth to collect user encounters and to transmit distress signal. The core is the trust model, which is responsible for classifying and extracting the level of trust based on mobile user encounters and other machine sensed vicinity data. User interfaces provide graphical input capability to record victim response to a distress signal message. Finally, the protocol and its adaptation module are used along with message to send the distress signal.

## VIII. TEST BED IMPLEMENTATION ANALYSIS

In this section, first we evaluate our test-bed results for Bluetooth performance using Nokia N810. We find that an average of 15-20 seconds is a good estimate for sending a distress signal to trusted nodes. Then, we analyzed over ten years of on-campus crime incidences against the density distribution of on-campus mobile users. Here, we found most of the incidences happen when mobile users are active. This positive correlation of 55% shows one good thing, incidences can be averted if a proper coordination and trust in the network exists.

### *A. Bluetooth Evaluation:*

Since Bluetooth is a primary mode for distress signaling, it is very important to first evaluate its performance and effectiveness to communicate the message. We used Nokia N810 to measure the Bluetooth performance on scanning and connection time, delivery and message size.

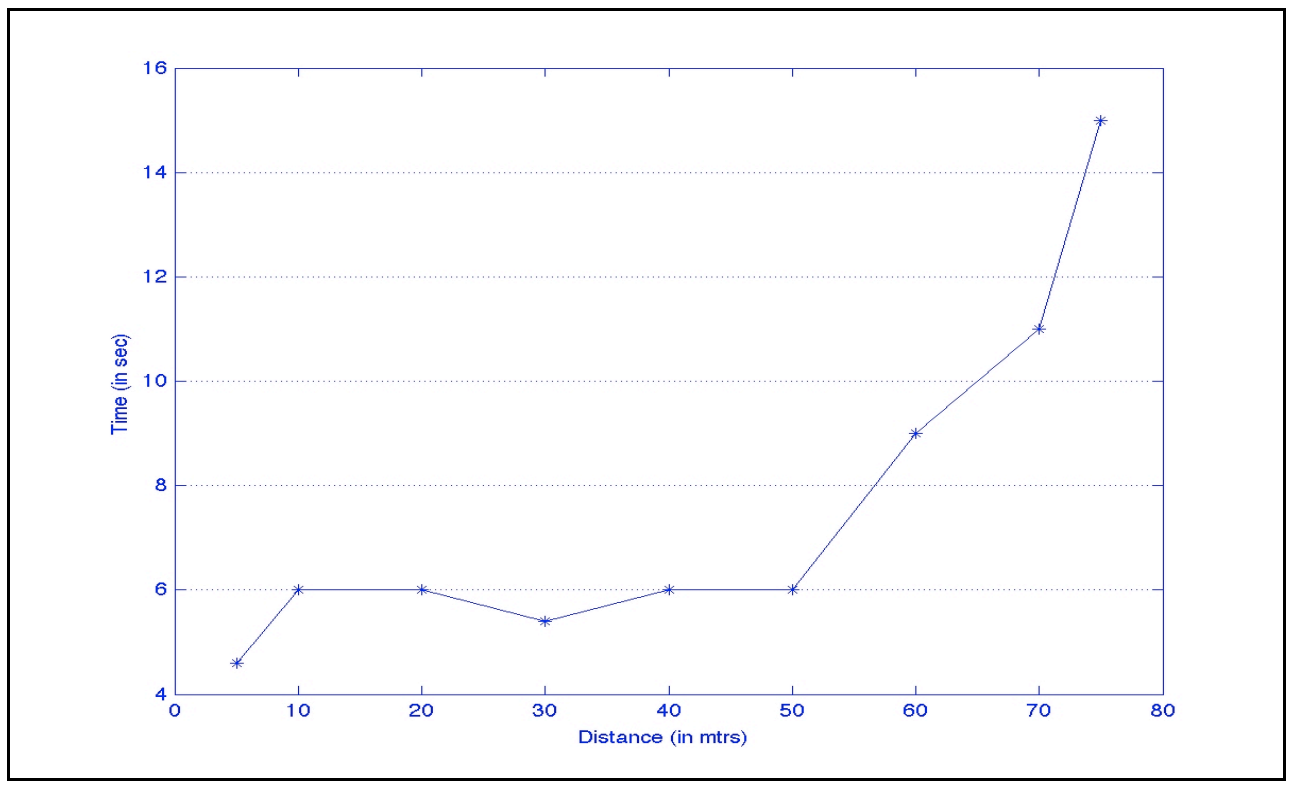

**Figure 5: As shown is the Bluetooth Scanning and Connection time, which varies with distance. We found that a good communication channel can be set in a range up to ~50 meters.**

To ensure the quickest level of communication, we optimized the Bluetooth capability of these devices. The result for average scanning and connection time is shown in Figure 5. The scan time of six to ten seconds is optimal to find trusted nodes within 50 meters of radius. As the emergency is relaxed, we can spend extra time in scanning and tracing the trusted nodes. We also define an efficient and useful Message format. We analyzed the connection and transfer time taken for a One Hop Transfer of message size of 184 bytes. As shown in Figure 6, we found that low transfer times range between 0-60 meters. These results show great promise to use Bluetooth communication in designing rescue applications.

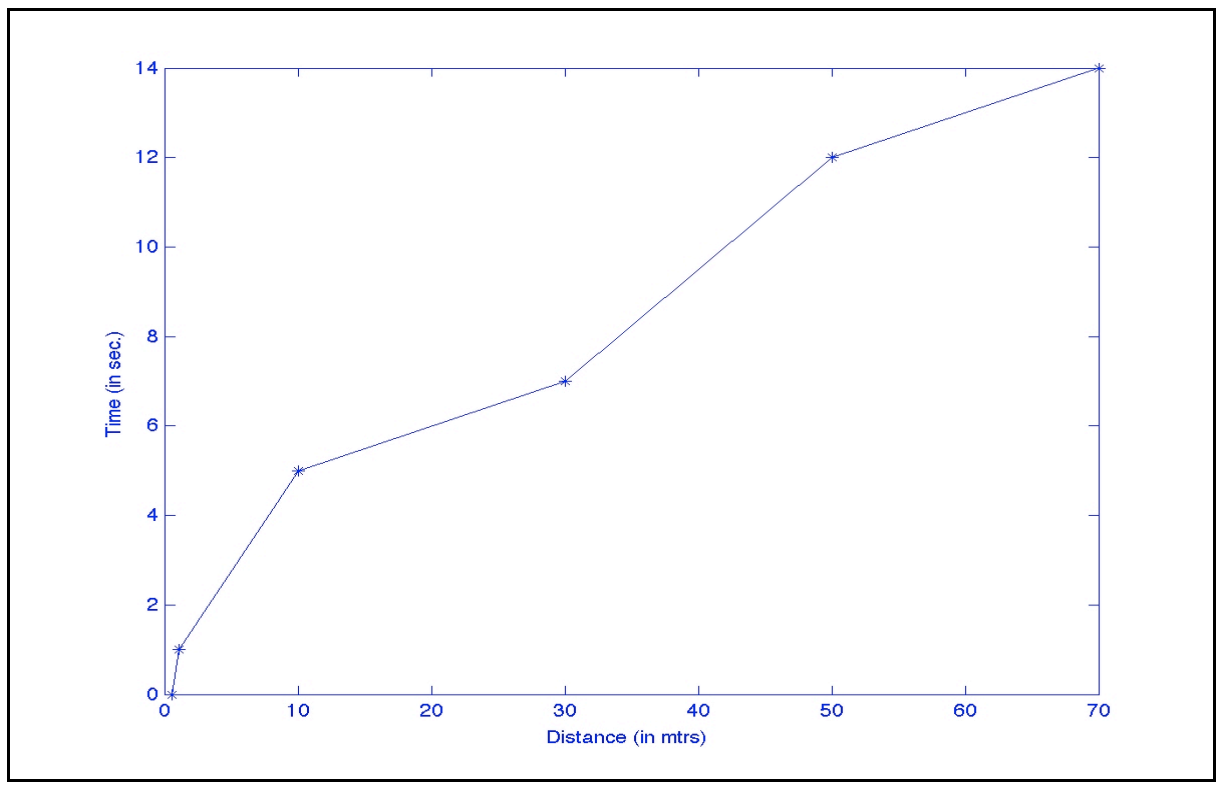

**Figure 6: Connection and Transfer time for Bluetooth**

### *B. Crime Statistics and Mobile User Density:*

Next, we analyzed the past ten years of the crime log statistics of University of Florida to understand the spatio-temporal distribution of the incidences happened on-campus and also the density distribution of active mobile users. As shown in Figure 7, the crime statistics are high during the midnight and then increases again as the daytime progresses. There is a positive correlation between the incidences and the number of active mobile users. Thus, these incidences can be very well averted given proper preparedness exists for the mobile users.

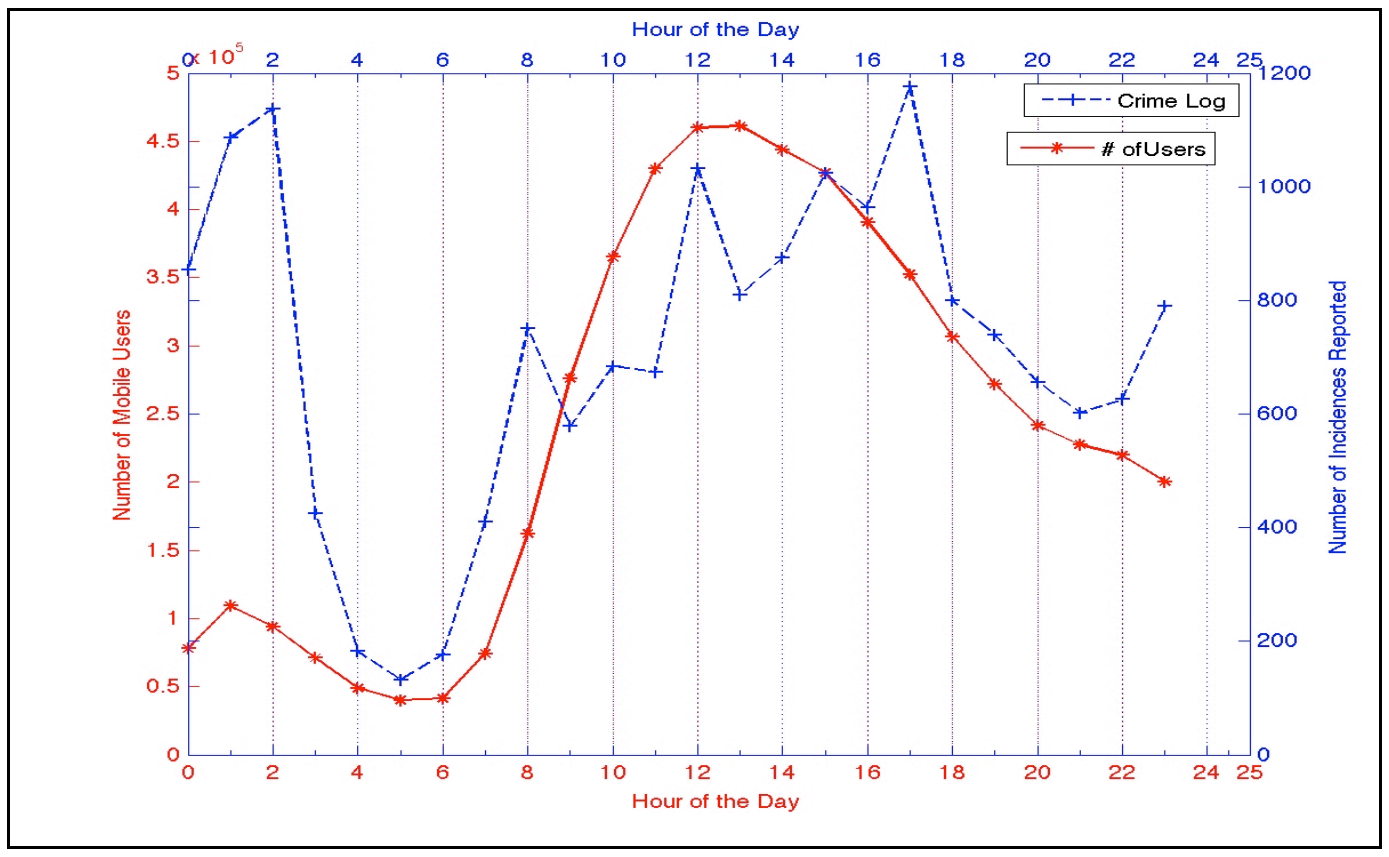

**Figure 7: A graph showing crime log and density distribution of on-campus mobile users. There is a 55% positive correlation between incidences and number of mobile users.**

## IX. CONCLUSION

In this paper, we propose a novel method to utilize handheld device in emergency rescue and alert scenarios. We introduced SHIELD, architecture to establish spatio-temporal trust and cooperation among mobile users based on their collaborative social influence for use in localized emergency alerts. An important sub-system of SHIELD is the proximity-enabled trust generation. To generate it, we use of the number and duration of Bluetooth encounters among mobile users. Our analysis shows that a large number of encounters and high meeting duration occurs among users who know each other very well. Then, we introduce a context-aware adaptive protocol that is both energy efficient and social aware for signaling distress message. Finally, we provided a proof-of-concept implementation for the use of Bluetooth as a viable communication medium. Our statistical analysis reveals a positive correlation between on-campus crime incidences and density distribution of users. The results indicate a need for such a system based on mutual trust and cooperation to avert incidences and help controlled flow of information during alerts. In the future, we plan to deploy the application prototype on iPhones and other handheld devices and provide a test-bed driven by real mobility measurements and routing protocols. We hope that SHIELD will augment the current safety infrastructure and for its deployment to help make a safer environment in schools and universities campuses, among others.

## ACKNOWLEDGMENT

We are thankful to the police department of University of Florida for providing us the vital crime log statistics.